\documentclass{elsart}
\usepackage{amssymb}
\usepackage{color}
\usepackage{url}
\usepackage{ifthen,graphics,epsfig}

\def\ko{K^0}

\def\ab{\bar{\alpha}}
\def\be{\begin{equation}}
\def\ee{\end{equation}}
\def\bea{\begin{eqnarray}}
\def\eea{\end{eqnarray}}

\newcommand{\kl}{\mbox{$K_L$}}
\newcommand{\ks}{\mbox{$K_S$}}

\newcommand{\eV}{{e\kern-.07em V}}

\def\figdir{} 

\def\mev{\hbox{ MeV}}
\def\kev{\hbox{ keV}}

\def \phidec {{\phi \to   K_S K_L }}
\def \ksdec  {{ K_S \to \pi^+ \pi^-}}

\makeatletter
\newdimen\z@ \z@=0pt 
\newskip\z@skip \z@skip=0pt plus0pt minus0pt
\def\m@th{\mathsurround=\z@}
\def\ialign{\everycr{}\tabskip\z@skip\halign} 
\def\eqalign#1{\null\,\vcenter{\openup\jot\m@th
  \ialign{\strut\hfil$\displaystyle{##}$&$\displaystyle{{}##}$\hfil
      \crcr#1\crcr}}\,}
\makeatother

\newcommand{\aff}[2]{Dipartimento di Fisica dell'Universit\`a #1 e Sezione INFN, #2, Italy.}
\newcommand{\affd}[1]{Dipartimento di Fisica dell'Universit\`a e Sezione INFN, #1, Italy.}

\begin{document}
\begin{frontmatter}
\title{Precise measurements of the  $\eta$ and the neutral kaon meson masses  with the KLOE detector.}
\collab{The KLOE Collaboration}

\author[Na]{F.~Ambrosino},
\author[Frascati]{A.~Antonelli},
\author[Frascati]{M.~Antonelli},
\author[Frascati]{F.~Archilli},
\author[Roma3]{C.~Bacci},
\author[Karlsruhe]{P.~Beltrame},
\author[Frascati]{G.~Bencivenni},
\author[Frascati]{S.~Bertolucci},
\author[Roma1]{C.~Bini},
\author[Frascati]{C.~Bloise},
\author[Roma3]{S.~Bocchetta},
\author[Frascati]{F.~Bossi},
\author[Roma3]{P.~Branchini},
\author[Roma1]{R.~Caloi},
\author[Frascati]{P.~Campana},
\author[Frascati]{G.~Capon},
\author[Frascati]{T.~Capussela},
\author[Roma3]{F.~Ceradini},
\author[Frascati]{S.~Chi},
\author[Na]{G.~Chiefari},
\author[Frascati]{P.~Ciambrone},
\author[Frascati]{E.~De~Lucia},
\author[Roma1]{A.~De~Santis},
\author[Frascati]{P.~De~Simone},
\author[Roma1]{G.~De~Zorzi},
\author[Karlsruhe]{A.~Denig},
\author[Roma1]{A.~Di~Domenico},
\author[Na]{C.~Di~Donato},
\author[Roma3]{B.~Di~Micco}
\footnote{Corresponding author: B.~Di Micco, e-mail dimicco@fis.uniroma3.it, M.~Dreucci, e-mail marco.dreucci@lnf.infn.it},
\author[Na]{A.~Doria},
\author[Frascati]{M.~Dreucci}\footnotemark[1],
\author[Frascati]{G.~Felici},
\author[Frascati]{A.~Ferrari},
\author[Frascati]{M.~L.~Ferrer},
\author[Roma1]{S.~Fiore},
\author[Frascati]{C.~Forti},
\author[Roma1]{P.~Franzini},
\author[Frascati]{C.~Gatti},
\author[Roma1]{P.~Gauzzi},
\author[Frascati]{S.~Giovannella},
\author[Lecce]{E.~Gorini},
\author[Roma3]{E.~Graziani},
\author[Karlsruhe]{W.~Kluge},
\author[Moscow]{V.~Kulikov},
\author[Roma1]{F.~Lacava},
\author[Frascati]{G.~Lanfranchi},
\author[Frascati,StonyBrook]{J.~Lee-Franzini},
\author[Karlsruhe]{D.~Leone},
\author[Frascati]{M.~Martini},
\author[Na]{P.~Massarotti},
\author[Frascati]{W.~Mei},
\author[Na]{S.Meola},
\author[Frascati]{S.~Miscetti},
\author[Frascati]{M.~Moulson},
\author[Frascati]{S.~M\"uller},
\author[Frascati]{F.~Murtas},
\author[Na]{M.~Napolitano},
\author[Roma3]{F.~Nguyen},
\author[Frascati]{M.~Palutan},
\author[Roma1]{E.~Pasqualucci},
\author[Roma3]{A.~Passeri},
\author[Frascati,Energ]{V.~Patera},
\author[Na]{F.~Perfetto},
\author[Lecce]{M.~Primavera},
\author[Frascati]{P.~Santangelo},
\author[Na]{G.~Saracino},
\author[Frascati]{B.~Sciascia},
\author[Frascati,Energ]{A.~Sciubba},
\author[Frascati]{A.~Sibidanov},
\author[Frascati]{T.~Spadaro},
\author[Roma1]{M.~Testa},
\author[Roma3]{L.~Tortora},
\author[Roma1]{P.~Valente},
\author[Frascati]{G.~Venanzoni},
\author[Frascati]{R.Versaci},
\author[Frascati,Beijing]{G.~Xu}

\clearpage
\address[Frascati]{Laboratori Nazionali di Frascati dell'INFN,
Frascati, Italy.}
\address[Karlsruhe]{Institut f\"ur Experimentelle Kernphysik,
Universit\"at Karlsruhe, Germany.}
\address[Lecce]{\affd{Lecce}}
\address[Na]{Dipartimento di Scienze Fisiche dell'Universit\`a
``Federico II'' e Sezione INFN,
Napoli, Italy}
\address[Energ]{Dipartimento di Energetica dell'Universit\`a
``La Sapienza'', Roma, Italy.}
\address[Roma1]{\aff{``La Sapienza''}{Roma}}
\address[Roma3]{\aff{``Roma Tre''}{Roma}}
\address[StonyBrook]{Physics Department, State University of New
York at Stony Brook, USA.}
\address[Beijing]{Permanent address: Institute of High Energy
Physics, CAS,  Beijing, China.}
\address[Moscow]{Permanent address: Institute for Theoretical
and Experimental Physics, Moscow, Russia.}

\begin{abstract}
We present precise measurements of the $\eta$ and $\ko$ masses using the processes
$\phi \to \eta \gamma$, $\eta \to \gamma\gamma$ and $\phidec$, $\ksdec$.
The $K^0$ mass measurement, $M_K=497.583\pm0.005_{\mathrm{stat}} \pm 0.020_{\mathrm{syst}}$ MeV, is in acceptable agreement with the previous measurements
but is more accurate.
We find $m_{\eta} = 547.874 \pm 0.007_{\mathrm{stat}} \pm 0.031_{\mathrm{syst}}$ MeV. Our value is the most accurate to date and is  
in agreement with two recent measurements based on $\eta$ decays, but  is
inconsistent, by about 10$\sigma$, with a measurement of
comparable precision based on $\eta$ production at threshold.

\end{abstract}

\end{frontmatter}
\maketitle

\def\ifm#1{\relax\ifmmode#1\else$#1$\fi}    
\def\ff{$\phi$--factory}  
\def\DAF{DA\char8NE}  
\def\f{\ifm{\phi}}   
\def\gam{\ifm{\gamma}}  
\def\epm{\ifm{e^+e^-}}  
\def\to{\ifm{\rightarrow}} 
\def\deg{\ifm{^\circ}}  
\def\ks{\ifm{K_S}}  
\def\kl{\ifm{K_L}}  
\def\dt{ \ifm{{\rm d}t} } 
\def\ab{\ifm{\sim}}
\def\x{\ifm{\times}}  
\def\up#1{$^{#1}$}  
\def\dn#1{$_{#1}$}   
\def\pic{\ifm{\pi^+\pi^-}}

\section{Introduction}

The $\eta$-meson mass has changed four times in the past 40 years while the measuring accuracy
never was better than 0.15 MeV till 2002. In 2005 the GEM experiment
using the reaction $d + p \to \eta\ ^{3}\mathrm{He}$ at threshold found $m_{\eta}
= (547.311 \pm 0.028_{\rm stat} \pm 0.032_{\rm syst})$ MeV~\cite{etamassGEM}, while in 2002 the NA48 collaboration 
using the decay $\eta \to
\pi^{0} \pi^{0} \pi^{0}$ found $m_{\eta} =  (547.843 \pm 0.030_{\rm stat} \pm
0.041_{\rm syst})$ MeV~\cite{etamassNA48}. The two results above differ by about
eight standard deviations. Preliminary KLOE results \cite{Cracovia} 
$m_{\eta} =  (547.822 \pm 0.005_{\mathrm{stat}} \pm0.069_{\mathrm{syst}})$ confirm the disagreement.
Recently the CLEO-c collaboration found
$m_{\eta} =  (547.785 \pm 0.017_{\rm stat}\pm 0.057_{\rm syst})$ MeV~\cite{etamassCLEO}
 using $\psi(2S) \to \eta J/\psi$ decays and combining different decay
modes of the $\eta$.

For the $K^0$ mass there is good  agreement   
between the Novosibirsk (1995) and CERN(2002) measurements that have a precision of $\sim 30$ keV \cite{k0CMD2,k0NA48}. 
Our measurement, similar to that of Novorsibisk with rather increased statistics, is based on the knowledge of the $\phi$ meson mass which
is known to 20 ppm from the Novosibirsk measurement employing the g-2 depolarizing resonance method. The $\phi$ mass  is also the basis for the $\eta$ mass measurement which relies on a precise determination of the collision center of mass energy, $W$ in the following. 
$W$ is determined run by run using $e^+ e^- \to e^+ e^-$ events ($\sim 40,000$ for each run), while the absolute momentum scale is obtained from the 
$e^+ e^- \to \phi \to K_S K_L$ cross section as a function of $W$.


\section{The KLOE experiment.}

KLOE operates at \DAF, the \ff\  \epm\ collider running  at a center of mass $W$ equal to the \f-meson mass. Positrons and electrons collide
at an angle of $\pi-0.025$ rad.
The KLOE detector consists of a 4 m diameter, 3.2m length drift chamber,
DC~\cite{KLOE-DC}, surrounded by a lead/scintillating-fiber
sampling calorimeter, EMC~\cite{KLOE-CAL}, both immersed in a
axial magnetic field of 0.52 T with the axis parallel to the
bisectrix of the two beam lines. The transverse momentum resolution for charged particles is $\delta
p_{\perp}/ p_{\perp} \simeq 0.4\%$. The EMC consists of a barrel and two end caps. Energy deposits in the EMC are reconstructed 
in the calorimeter with energy and time resolutions
$\sigma_{E}/E$ = 0.057/$\sqrt{E\ \mathrm{(GeV)}}$, $\sigma_{t}$ = 54 ps /
$\sqrt{E\ \mathrm{(GeV)}}$ in quadrature with 140 ps. The centroid of showers is
measured with resolution $\sigma_{\ell}$ = 1 cm/$\sqrt{E\ \mathrm{(GeV)}}$
in the coordinate parallel to the fibers and 1 cm in the
transverse coordinate.

 For a photon coming from the IP the
angular resolution is $\sigma \sim 1 \mathrm{cm} /200 \mathrm{cm} \sim 5 \, \mathrm{mrad}$.
Close-by energy deposits are combined into ``clusters''.
A prompt photon is defined as a cluster with $|t_{clu} -
r_{clu}/c| < 5\sigma_{t}$ ($t_{clu}$ is the arrival time measured at the EMC,
$r_{clu}$ is the distance from the $e^+ e^-$ interaction point and $c$ is
the velocity of light) not associated to a charged particle. For
this latter, we require the distance between the centroid of the cluster
and the extrapolation of any track reaching the calorimeter to be larger than three times
the cluster position resolution.

Only calorimeter signals are used to trigger~\cite{KLOE-TRIG} events for these analyses. We require at least two energy
deposits above threshold ($E > 50$ MeV in the barrel and $E > 150$
MeV in the end-cap). The trigger has a large time jitter with
respect to the event time but is synchronized with the collider radio frequency with an accuracy of
50 ps. The time of the bunch crossing producing an event is
determined off-line during event reconstruction.

The large cross section for \epm\to\f, \ab 3 $\mu$b and for elastic \epm\ scattering, Bhabha scattering, allows KLOE to collect large number of events, some 450,000 per hour. KLOE takes advantage of these events to maintain a running calibration of time and  energy scales of its calorimeter, of the momentum and position resolution of the drift chamber, of the machine energy and beams crossing angle, and therefore of the center of mass motion,  of the mean position of the interaction point and of the detector alignment.

\section{Calibration of c.m. energy.}

 The center-of-mass energy, $W$, has been measured for 
 each run by fitting  the $e^+ e^-$ invariant-mass distribution 
 for Bhabha events to a Monte Carlo generated function, including radiative effects.

 Initial state radiation (ISR), 
 where one or both initial colliding 
 particles radiate a photon before interacting, affects the $e^+ e^-$ collision
 center-of-mass energy $W$ and therefore the final state invariant mass. 
 ISR, which is mostly collinear to the beam, is in general not detected.
MC Bhabha events were generated   using the BABAYAGA event generator~\cite{babayaga},
 which accounts for both  final and  initial state radiation.

 An example of this fit is shown in Fig.~\ref{fit}.
\begin{figure}[h]
  \begin{center}
    \mbox{\epsfig{file=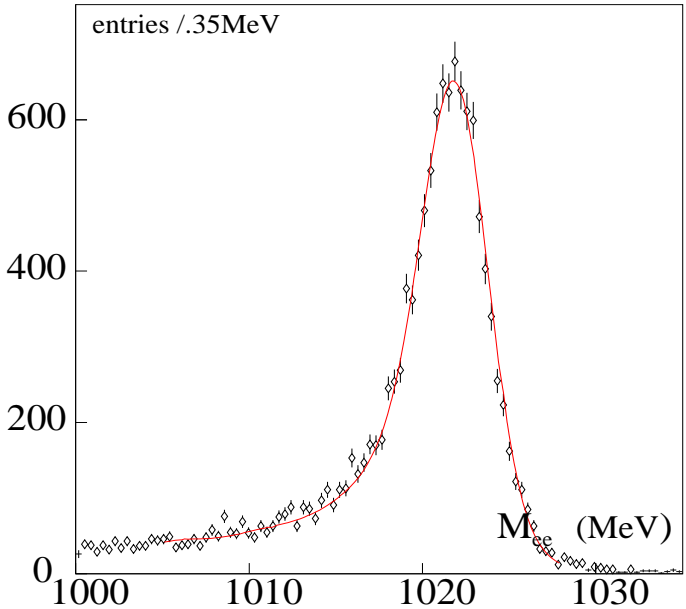,width=9cm,height=8cm}}
    \caption{Fit to reconstructed $e^+ e^-$ invariant-mass 
             distribution for Bhabha events, for a run with $W$=1021.7~MeV} 
    \label{fit}
  \end{center}
\end{figure}
 In a typical run,  an integrated luminosity  of 50 ${\rm nb}^{-1}$
 is collected and $W$
 is measured with a statistical accuracy of $\sim 3$ keV. 
 The stability of the momentum calibration has been studied 
 measuring the two-pion invariant mass in $\ksdec$ decay. It
 is found to be stable to within 10 keV in the analyzed runs.
 
 The center-of-mass energy scale has been calibrated
 by obtaining the $\phi$ mass from a fit to the cross section measurements
 for the process $e^+ e^- \to $$\phidec$.
 The cross section is measured at different values of $W$ around 
 $M_{\phi}$ by counting the number of $\ksdec$ events,  correcting  
 for selection efficiency, background and the $\beta^3_K$ factor of the $K_S K_L$ pair, and normalizing to the integrated 
 luminosity. 
 $\ksdec$ events are selected by requiring two 
 tracks with opposite charge that form a vertex 
 within a cylinder of 5 cm radius and 10 cm length centered on the
 interaction point. The invariant mass computed from the two pion tracks 
 is required to be within 20 MeV  of the nominal neutral kaon mass.
 The luminosity is measured by using very-large-angle ($>55^0$) Bhabha events\cite{KLOE-LUM}.
 

 The measured cross section is fitted to a theoretical function~\cite{CMD2par}
 that depends on the $\phi$ parameters, takes into account the  
 effect of ISR, and includes the interference with the $\rho(770)$ and
 the $\omega(782)$ mesons.
 The $\phi$ mass, total width, and peak cross section are the
 only free parameters of the fit, the $\rho(770)$ and  the $\omega(782)$
 parameters being fixed.
 The results of the fit to the data are shown in Fig.~\ref{line}. The 
 fitted $\phi$ mass is $M_{\phi} = 1019.329 \pm 0.011 \mev$,  
 to be compared with 
 $ M_\phi= 1019.483 \pm 0.011 \pm 0.025 \mev$, measured by CMD-2
 at VEPP-2M ~\cite{CMD2}.
 The ratio of these two values is used to fix the overall energy scale.
 The correction factor $M_{\phi}^{CMD}$/$M_{\phi}^{KLOE} $ is $1.00015$, 
 corresponding to a shift in the value of $W$ of \ab150 keV.
\begin{figure}[h]
  \begin{center}
    \mbox{\epsfig{file=\figdir 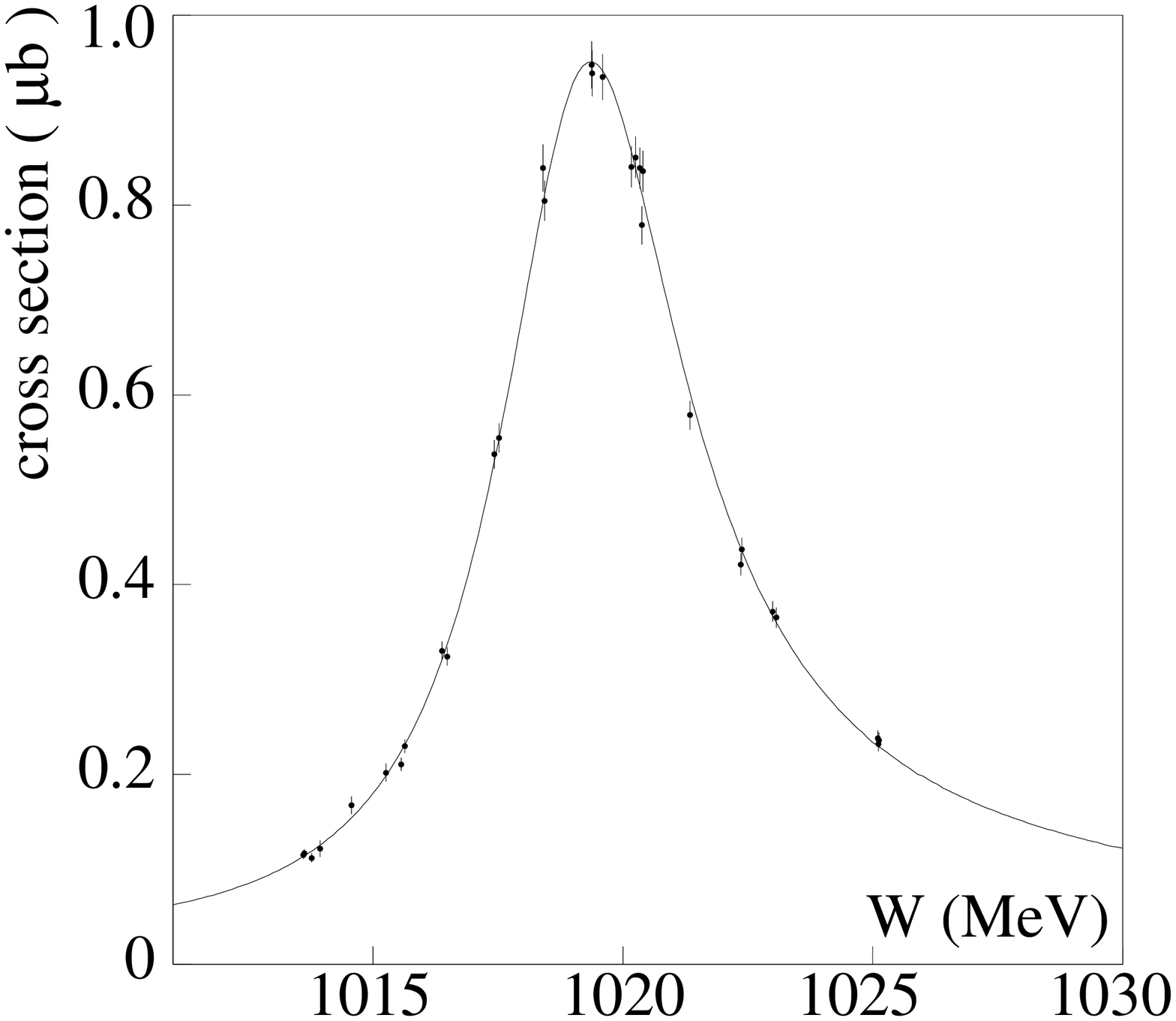,width=9cm,height=8cm}}
    \mbox{\epsfig{file=\figdir 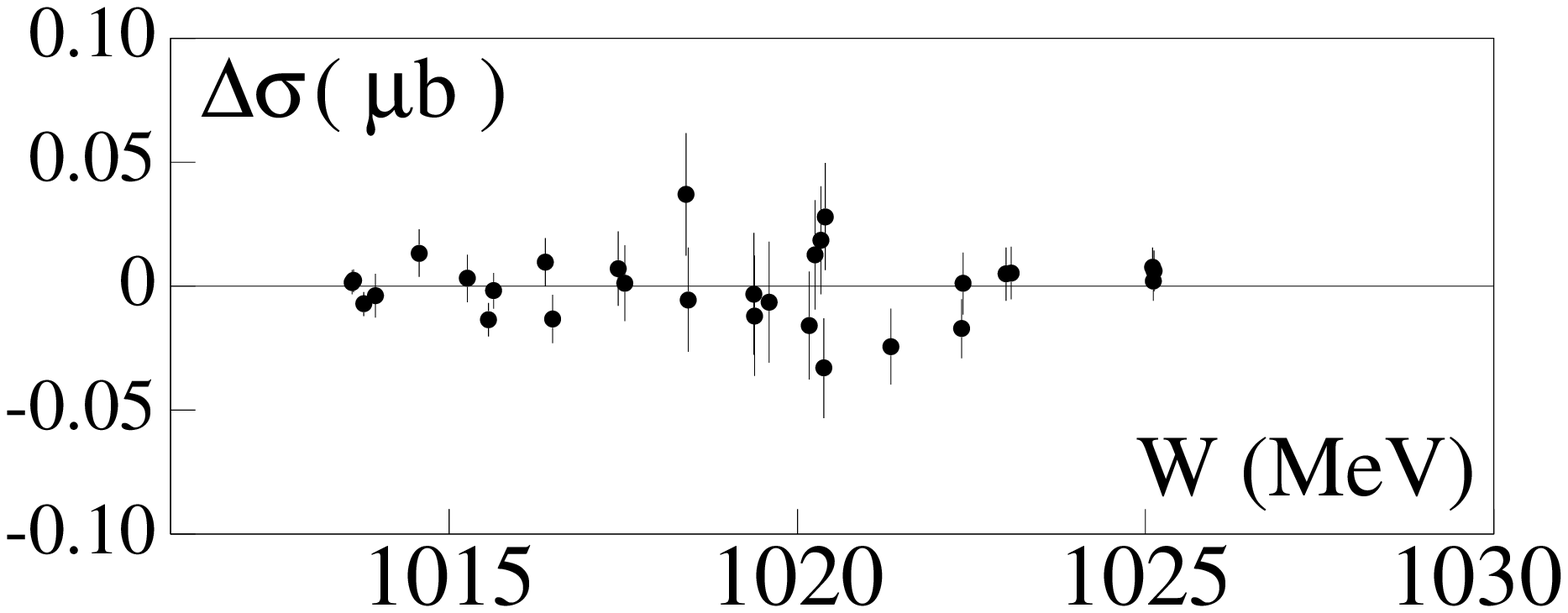,width=9cm,height=3.5cm}}
    \caption {Top: Cross section for $e^+ e^-\to$$\phidec$ as a function
              of the center-of-mass energy. 
              The solid line represents the fit to the data. Bottom:  Fit residuals.} 
    \label{line}  
  \end{center}
\end{figure}

\section{Measurement of the neutral kaon mass.}
The events $\phi \to K_S K_L$ offer a unique possibility to obtain a precise value of the neutral kaon mass. To obtain a crude estimate 
of the resolution and explain the method used we  observe that if the $\phi$-meson is at rest
the kaon mass can be extracted from the kaon momentum using the relation:
\begin{equation} \label{eq:kmasssimple}
m_{K} = \sqrt{\frac{m_{\phi}^2}{4} - p_{K}^2};  \qquad \frac{\Delta m_{K}}{m_{K}} \simeq \frac{p_K^2}{m_{K}^2}\frac{\Delta p_K}{p_K} \sim \beta^2 \frac{\Delta p_K}{p_K}
\end{equation}
Since $p_K \simeq 110$ MeV, measuring it at  1\% level, well within the KLOE capability, results in a measurement of the $K^0$ mass better than $0.1$\%. 
50,000 events are enough to reach a statistical accuracy of about 1 keV.

$\phi$ mesons are produced with a momentum along the $x$ axis, $p_{\phi} = 12.5$ MeV at DAFNE.
From the measured momenta of the two pions from $K_S \to \pi^+ \pi^-$, we  measure the $K_S$ momentum. The $K_L$ momentum is given by 
$\vec{p}_{K_L} = \vec{p}_{\phi} - \vec{p}_{K_S}$, where $\vec{p}_{\phi}$ is the average $\phi$ momentum  
 measured with Bhabha events collected in the same runs.
The center of mass energy of the $K_S K_L$   pair ($W_KK$) is related to the kaon mass $M_K$, according to:
\[
W_{KK}(M_K) =\sqrt{2M^2_K+2E_{K_S}E_{K_L}-2\vec{p}_{K_S}\cdot\vec{p}_{K_L}} 
\]
with
\[
E_{K_S} = \sqrt{p_{K_S}^2+M^2_K} \qquad \qquad E_{K_L} = \sqrt{p^2_{K_L}+M^2_K}.
\] 
On the other hand the  collision center of mass energy $W$ is computed from Bhabha events as described above.

 Corrections due to ISR have to be taken into account
 when relating  $W$ to   $W_{KK}$.
 The correction function $f_K(W)$ 
 has been evaluated using a full detector simulation where
 the radiation from both beams has been implemented, and $W_{KK}$
 is reconstructed as in the data.
 The expression of the radiator function 
 has been taken from Ref.~\cite{nicrosini}, including 
 $\mathcal{O}(\alpha^2)$ corrections.
 The correction $|1-f_K(W)|$ is very small below the resonance: 
about 40 $\kev$. 
 It  increases up to {\color{red}} about 100 keV when $W$ is above 
 the $\phi$ mass. In this region  radiative return begins to be 
 important. The neutral kaon mass is then obtained solving the equation:
 \begin{equation}
  W = f_K(W) \cdot W_{KK}(M_K) \label{eq:kaonmass}
 \end{equation}
 The single event mass resolution is  about 430 $\kev$. Contributions to the mass resolution are: experimental resolution 
 about 370 keV, beam energy spread about 220 keV, as measured by KLOE 
 {\color{blue}(} in agreement with machine theory {\color{blue})}, and ISR about 100 keV. The kaon mass distribution for a single run is shown in Fig.~\ref{fig:kpeak} together with a gaussian fit to the distribution.
\begin{figure}
\begin{center}
\includegraphics[width=0.7\textwidth]{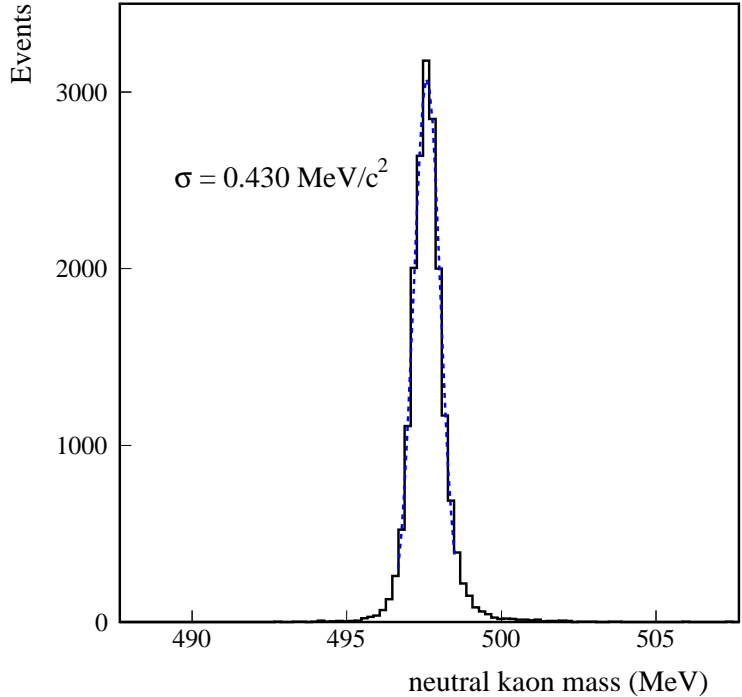}
\end{center}
\caption{$M_K$ distribution evaluated in a single run of $\sim$ 16000 events from eq. \ref{eq:kaonmass}, the dotted line is a gaussian fit to the peak.} \label{fig:kpeak}
\end{figure}

The source of systematic errors considered for this measurement are: 
\begin{enumerate}
\item the momentum calibration;
\item the theoretical uncertainty on the radiator function $f_K(W)$;
\item the absolute calibration of the beam energy.
\end{enumerate}

 The systematic error due to the momentum miscalibration has been evaluated by
changing the momentum scale in computing the pion momenta. 
A momentum miscalibration, $\delta p/p$ , translates to a miscalibration on  $\delta M_{K} /M_{K}$= 0.06 $\delta p/p$, 
in agreement with the qualitative calculation made above. 
The momentum scale is obtained by using several processes  covering a wide momentum range 50 to 500 MeV 
($K_L \to \pi^+ \pi^- \pi^0, ~K_L \to \pi \ell \nu, ~\phi \to \pi^+ \pi^- \pi^0$), with a fractional 
accuracy below $2 \times 10^{-4}$, in agreement  with the estimate obtained using Bhabhas \cite{KLOE-MC}, resulting in a systematic error 
$\delta M_K$ of 6 keV.

The systematic error coming from theoretical uncertainty of the radiator function has been 
evaluated considering the contribution  from higher order terms in $\alpha$. 
The correction function $f_K (W )$ has been evaluated by excluding the constant term in the $\mathcal{O}(\alpha^2)$. 
The corresponding change in $f_K(W)$ is $1.3 \times 10^{-5}$ corresponding to a variation on $M_K$ of 7 keV. 
Further checks have been made by using the  function given in Ref. \cite{Kuraev}: no significant differences were observed.
Additional systematics come from the dependence of the measured mass from the $W$ value: 
we compare  the average of the measurements with data collected at $W<1020$ MeV with data at $W>1021$ MeV,
 where the value of $f_K(W)$ is more than a factor two larger.
 The difference between the two mass values is   $M_K(W<1020)-M_K(W>1021)=9\pm10$ keV, consistent with zero.

Other sources of systematics are due to the uncertainties on the W calibration, i.e., the statistic and systematic error on $M_{\phi}^{CMD-2}$ and on $M_{\phi}$ obtained from our fit. The
total contribution from these sources amounts to a mass uncertainty of 15 keV. Systematic uncertainties are treated as uncorrelated. The result is:
\begin{equation} \label{eq:EQK0}
M_K = 497.583 \pm 0.005_{\mathrm{stat}} \pm 0.020_{\mathrm{syst}} \quad \mathrm{MeV}.
\end{equation}

\section{The $\eta$ mass}
The decay  $\phi \to \eta \gamma$, for $\phi$-meson at rest,  
is a source of monochromatic $\eta$-mesons of $\sim$ 362.8 MeV momentum, recoiling against a photon of equal momentum. Detection of such a photon signals the presence of an $\eta$-meson. Photons from $\eta$\to\gam\gam\ have a  flat  spectrum in the range $147 < E_{\gamma} < 510$ MeV in the laboratory frame. 
In the laboratory, the opening angle of the two photons has a distribution peaked at its minimum value of 113\deg~. KLOE measures this  angle with an accuracy of \ab0.4\deg. The value of the minimum angle is a function of the $\eta$ mass and its measurement determines the mass with a resolution of 2 MeV, without energy measurements. In fact we do measure the photon energies. The  $\eta$-mass accuracy is however ultimately due to the accurate measurement of the photon angles. Together with the stability of the continuously calibrated detector and the very large sample of $\eta$-mesons collected we have been able to obtain a very accurate measurement of the $\eta$-mass \cite{nota}.


Events are
selected requiring at least three energy clusters in the barrel
calorimeter with polar angle $50^{\circ} < \theta_{\gamma} <
130^{\circ}$. A kinematic fit imposing energy-momentum
conservation is performed. The fitted photon energy resolution is vastly improved 
over the EMC measurement because of the good angular resolution.
 The kinematic fit uses the value of the
total energy, the $\phi$ transverse momentum and
the average value of the beam-beam interaction point; these values
are determined with good precision run by run by analyzing $e^{+}
e^{-} \to e^{+} e^{-}$ elastic scattering events.
Fig.~\ref{CHI2} shows the $\chi^{2}$ of the kinematic fit for the
data and for Monte Carlo~\cite{KLOE-MC} simulated signal events.
If more than three photons are selected, the combination with the
lowest $\chi^{2}$ is chosen. Events with  $\chi^{2} < 35$ are kept for the analysis.

\begin{figure}[htb]
    \centering
 \mbox{\epsfig{file=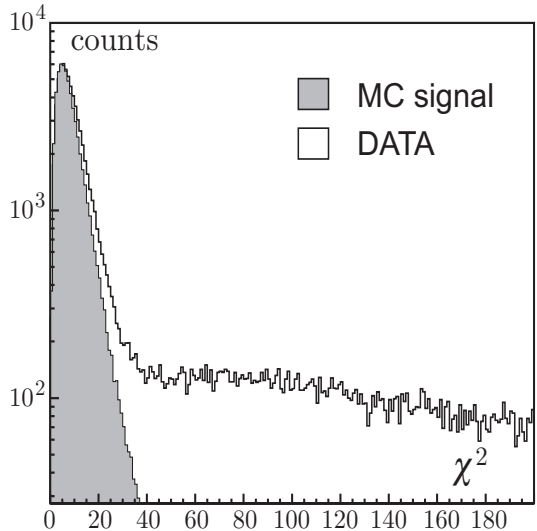,width=7cm,height=7cm}}
    \caption{Distribution of the $\chi^{2}$ of the
    kinematic fit: dashed area for the MC simulation of $\phi \to
    \pi^{0} \gamma$, $\phi \to \eta \gamma$ events; histogram for
    data. }
    \label{CHI2}
\end{figure}

Fig.~\ref{DALITZ} shows the $m^{2}_{\gamma_{2} \gamma_{3}}$,$m^{2}_{\gamma_{1}
\gamma_{2}}$  Dalitz plot population, with the energies ordered as $E_{\gamma_{1}} <
E_{\gamma_{2}} <  E_{\gamma_{3}}$. 
The $m^2_{\gamma_{1}
\gamma_{2}} \simeq m^2_{\pi^{0}}$,  $m^2_{\gamma_{1}
\gamma_{2}} \simeq m^2_{\eta}$ and  $m^2_{\gamma_1 \gamma_3} = m_{\eta}^2$ bands are clearly visible.
We apply a cut $m^{2}_{\gamma_{1} \gamma_{2}} +
m^{2}_{\gamma_{2} \gamma_{3}} \leq 0.73$ GeV$^2$, ``background-rejection cut'' in the following, shown by the line in
Fig.~\ref{DALITZ}.  Events below the line are retained for the analysis. The
resulting $m_{\gamma_{1} \gamma_{2}}$ distribution, for a data subsample, 
is shown in Fig.~\ref{MASS12},top. The $m(\gamma_1 \gamma_2)$ distribution in the 542.5 to 552.5 interval is fitted well
with a single gaussian with $\sigma = 2.0$ MeV as shown in
Fig.~\ref{MASS12},bottom. 

\begin{figure}[htb]
    \centering
 \mbox{\epsfig{file=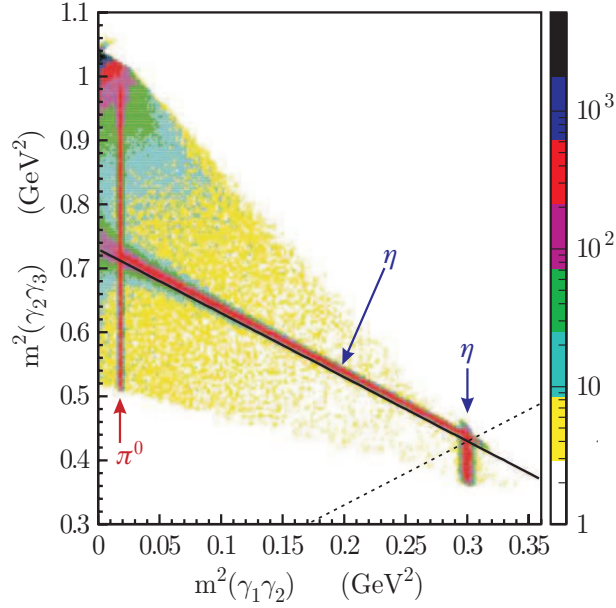,width=8cm,height=8cm}}
    \caption{Population in the
    $m^{2}_{\gamma_{2} \gamma_{3}}$, $m^{2}_{\gamma_{1}
    \gamma_{2}}$ plane. The photon energies are ordered as $E_{\gamma_{1}} <
    E{\gamma_{2}} <  E_{\gamma_{3}}$. The $\eta$ and $\pi^0$ signal are quite evident. Dashed line see text.}
    \label{DALITZ}
\end{figure}

\begin{figure}[htb]
    \centering
 \mbox{\epsfig{file=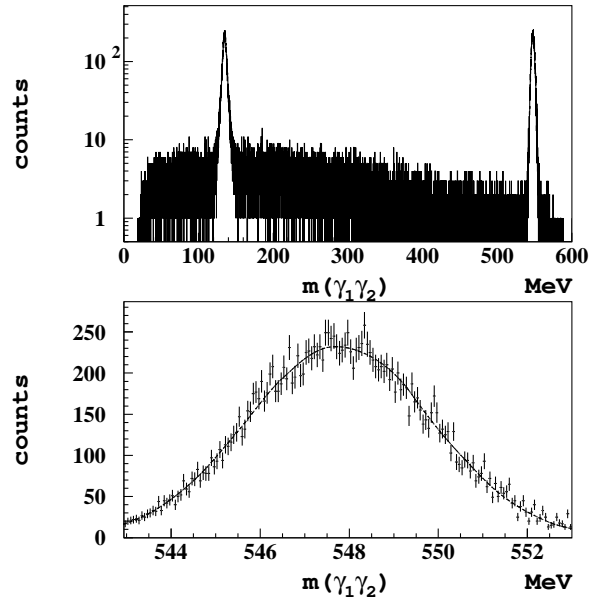,width=8cm,height=8cm}}
    \caption{Top: Distribution of the invariant mass
    $m({\gamma_{1} \gamma_{2}})$, for the events selected by the cut shown in Fig.\ref{DALITZ}
    Bottom:
    Distribution of the invariant mass $m({\gamma_{1} \gamma_{2}})$
    around the value of the $\eta$ mass and the gaussian fit. The result of the fit is $m_{\eta} = 547.777
\pm 0.016$ MeV with $\chi^{2}$/n.d.f = 168/161, CL=33\%.}
    \label{MASS12}
\end{figure}

To estimate systematic uncertainties we have studied the effects of the detector response and
alignment, event selection cuts, kinematic fit and beam energy
calibration that can influence our measurement. The values of the
systematic errors are summarized in Table~\ref{SYSTEMATIC}.

First, to check the effect of the $e^+ e^-$ interaction point and the alignment of the calorimeter relative to the drift
chamber, we have selected a high purity  sample of $e^{+} e^{-} \to \pi^{+}
\pi^{-} \gamma$ events \cite{hadroncross}. The average position of the interaction
point, determined run by run with $e^{+} e^{-} \to e^{+} e^{-}$
events, has been compared with the average position of reconstructed $\pi^{+} \pi^{-}$
vertex . The difference between the two values was computed
run by run and the rms of these points ($\sigma_{\mathrm{vtx}}$) was used to evaluate the
systematic error introduced in the kinematic fit by varying the IP position by $\pm 1 \sigma_{\mathrm{vtx}}$.
 To check for misalignments
between the calorimeter and the drift chamber, each pion track was
extrapolated to the calorimeter and compared with the centroid of
the cluster. A small correction of 1.1 mm along the vertical
coordinate, $y$, and of 2.0 mm along the longitudinal coordinate,
$z$, was applied. The rms ($\sigma_{\mathrm{displ}}$) of the
mean difference between the coordinates of the extrapolated point and
the cluster centroid was used to compute the systematic uncertainty on the $\eta$ mass by shifting the photon point of arrival at the EMC by $\sigma_{\mathrm{displ}}$.

The energy scale of the calorimeter response and its linearity are
checked using two different samples of $e^{+} e^{-} \to \pi^{+} \pi^{-} \gamma$ and
$e^{+} e^{-} \to e^{+} e^{-} \gamma$ events. The energy of the
photon, determined from the track momenta and the average value of
$W$, was compared with the calorimeter cluster energy. The calorimeter energy scale was calibrated to
better than 1\% and  the response is linear to better than
2\% in the range of interest. The systematic effect on the
two-photon invariant mass is 4 keV from the energy scale miscalibration
and  4 keV from the non-linearity.
The values above confirm that the mass measurement has little
sensitivity to the calorimeter energy response.

It is however important
to check the correctness of the position measurement in the
24 calorimeter modules of the barrel. We compute the two-photon invariant
mass for different orientations of the $\gamma_1 \gamma_2 \gamma_3$
plane. The rms width of the $\eta$ mass distribution was assumed as
systematic error: 10 keV and 15 keV respectively
for variations of the polar and azimuth angle of the normal to the
plane.

Systematic effects from event selection criteria were studied by varying
the $\chi^{2}$ cut (Fig.~\ref{CHI2}) and the background-rejection cut of
Fig.~\ref{DALITZ}. The first has no influence ($< 1$ keV)
on the result; the second introduces a systematic error of 17
keV determined by translating the solid line in Fig.~\ref{DALITZ} and rotating as 
shown by the dashed line in the same figure. 
The mass value is very sensitive to the center of mass energy of
the $\eta \gamma$ system used in the
kinematic fit.  Due to initial state radiation
emission (ISR) the available center of mass energy  is a bit lower than
the value computed from the nominal energy of the $e^+, e^-$  beams.
The effect has
been studied with a detailed Monte Carlo simulation of  the events
in the detector, and a shift of $\sim$100 keV
was found for the mass measurement. Since this correction is
relatively large, we have checked the MC correction due to ISR
emission also for runs taken at different values of $W$. 
The data were divided in eight energy bins; moreover, two
off-peak energy bins, centered at $W$ = 1017 and 1022 MeV,
were also analyzed in the same way as the $\phi$-peak data.
Fig.~\ref{COR-ROOTS}, top shows the shift of the mass evaluated by MC 
 as a function of $W$
together with the measured shift of the
$\eta$ mass respect to the value obtained at $W = 1019.6$ MeV.
Fig.~\ref{COR-ROOTS}, bottom shows the value of the mass corrected for the
ISR effect. The rms of these points is used as systematic error
(8 keV).

\begin{figure}[htb]
    \centering
 \includegraphics[width=0.7\textwidth]{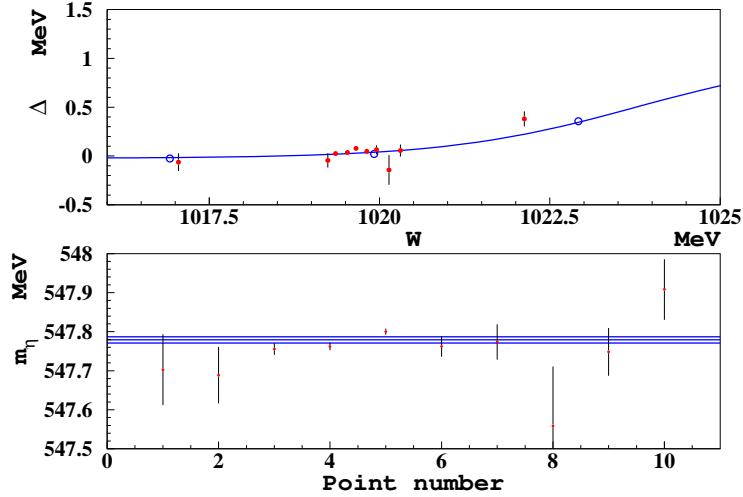}
    \caption{Top. Shift of the value of the $\eta$ mass due to ISR
    as a function of $W$. Open dots: MC,  full dots: data ($m_{\eta}(W)-m_{\eta}(1019.6)$). The
    line is a fit to the MC.
    Bottom. Value of the $\eta$ mass after the correction as a
    function of $W$. The band corresponds to $\pm 1 \sigma$ around the fitted value.}
    \label{COR-ROOTS}
\end{figure}

The value of the $\pi^{0}$ mass was measured with the same method
fitting the low mass region of Fig. \ref{MASS12} and the ratio $R = m_{\eta} / m_{\pi^{0}}$
was also determined. All systematic effects discussed above were also
evaluated for the mass of the $\pi^{0}$ and for the ratio $R$; the
corresponding values are listed in Table~\ref{SYSTEMATIC}.

\begin{table}[hbt]
   \begin{center}
   \begin{tabular}{|l|c|c|c|} \hline
   Systematic effect & $m_{\eta}$ (keV) &
   $m_{\pi^{0}}$ (keV) & R ($\times 10^{-5}$) \\
   \hline
   Vertex position              & 4  & 6 & 19 \\
   Calorimeter energy scale     & 4  & 1 & 6 \\
   Calorimeter non-linearity    & 4  & 11 & 31 \\
   $\theta$ angular uniformity  & 10 & 44 & 120 \\
   $\phi$ angular uniformity    & 15 & 12 & 37 \\
   $\chi^{2}$ cut               & $<$1  & 4 & 13 \\
   Background-rejection cut     & 17 & 4 & 18 \\
   ISR emission                 & 8  & 9 & 28 \\
   \hline
   Total                      & 27 & 49 & 136 \\
   \hline
   \end{tabular}
   \end{center}
   \caption{Systematic errors evaluated for $m_{\eta}$, $m_{\pi^{0}}$
   and the ratio $R = m_{\eta}/m_{\pi^{0}}$.}
\label{SYSTEMATIC}
\end{table}

Finally, the stability of the results as function of  running
conditions was checked by dividing the data set in eight different
periods and determining the values of $m_{\eta}$, $m_{\pi^{0}}$
and $R$ for each period. The results are shown in
Fig.~\ref{PERIODS} and in Table~\ref{RESULTS} together with their
statistical significance. Fit to a common value are good.

\begin{figure}[hbt]
    \centering
     \resizebox{4.3cm}{!}{\includegraphics{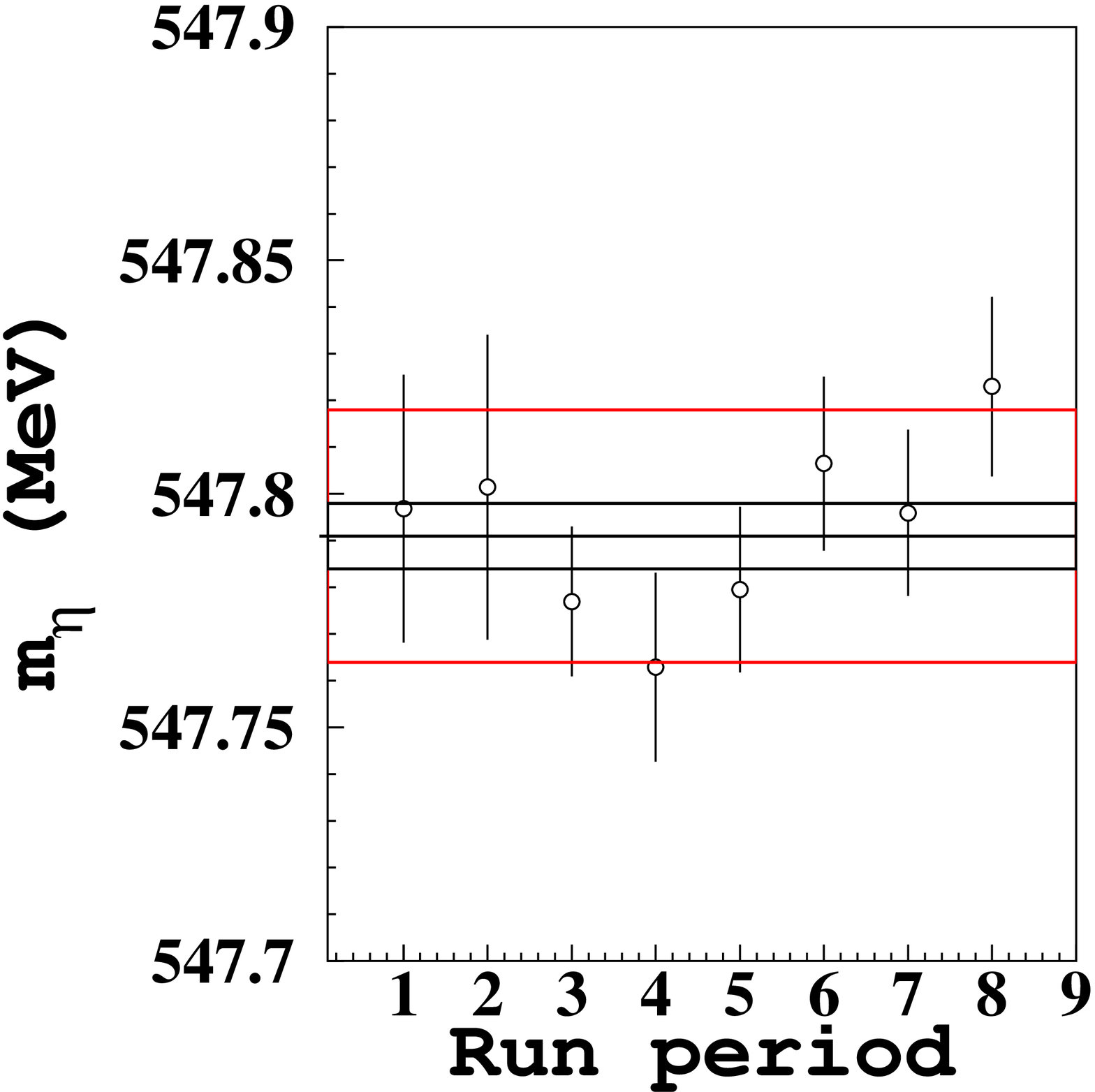}}
     \resizebox{4.3cm}{!}{\includegraphics{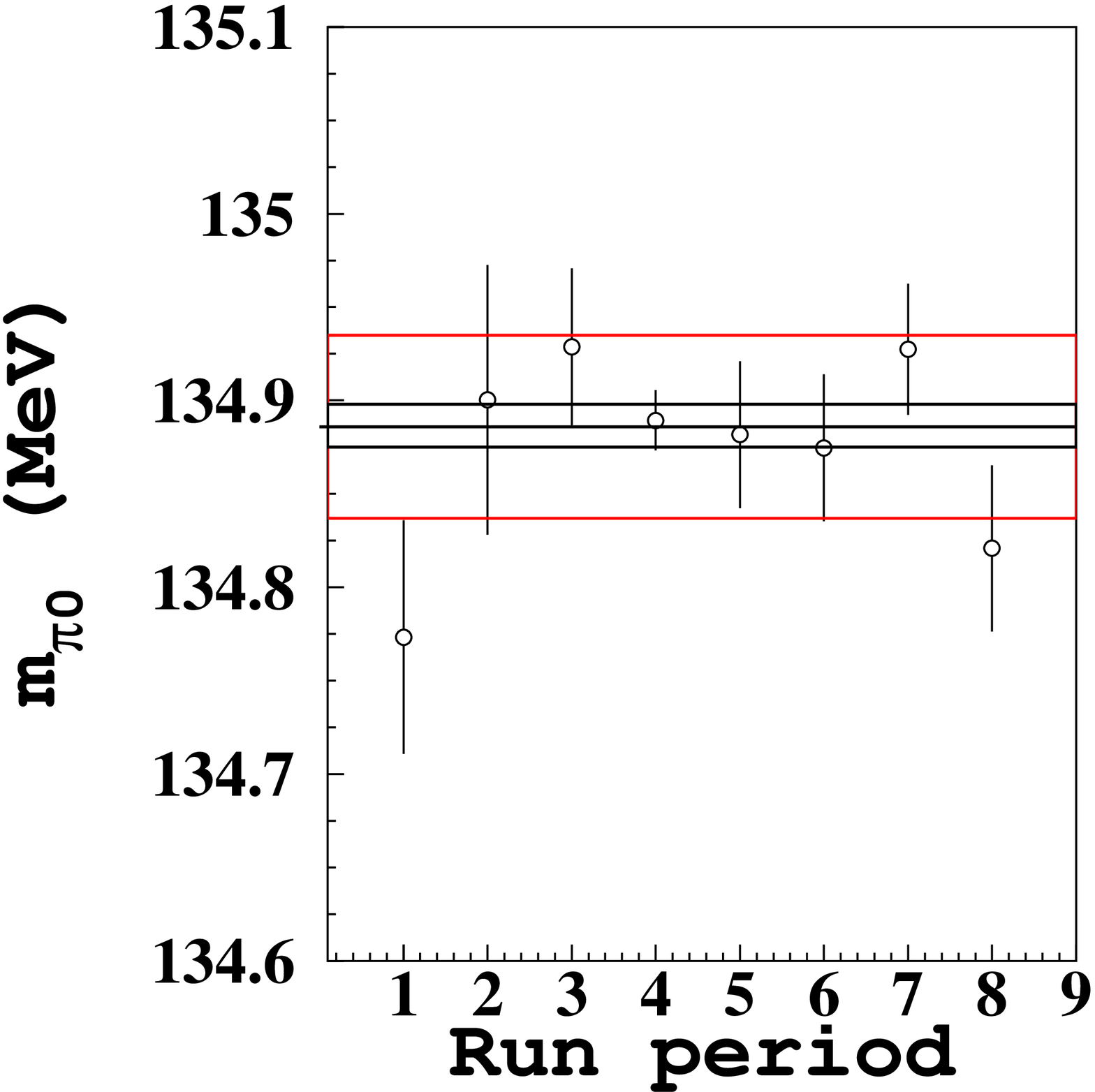}}
     \resizebox{4.3cm}{!}{\includegraphics{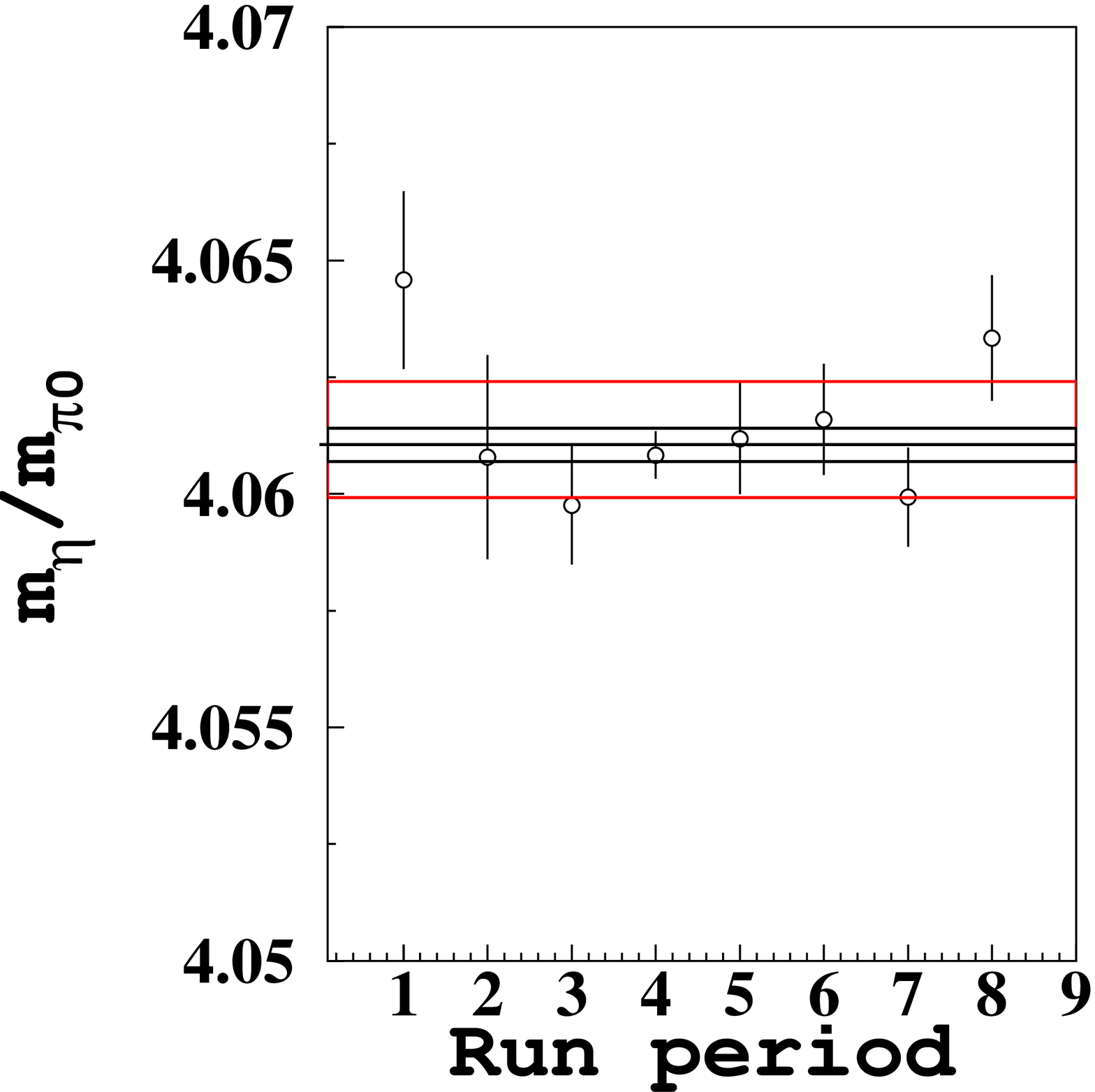}}
    \caption{Value of the $\eta$ mass, the $\pi^0$ mass and
of the ratio  measured in the eight run periods. The narrow (large) bands correspond to statistical (systematic) error.}
    \label{PERIODS}
\end{figure}

\begin{table}[htb]
   \begin{center}
   \begin{tabular}{|l|c|c|c|c|} \hline
    & Result of the fit & $\chi^{2}$/\mbox{d.o.f.} & C.L. \% \\
   \hline
   $m_{\eta}$ & 547,791 $\pm$ 7 (keV) & 6.9/7 & 44 \% \\
   $m_{\pi^{0}}$ & 134,886 $\pm$ 12 (keV) & 7.7/7 & 36 \%\\
   $R$ & 4.0610 $\pm$ 0.0004 & 8.9/7 & 26 \%\\
   \hline
   \end{tabular}
   \end{center}
   \caption{Result of the fit to the values of $m_{\eta}$,
   $m_{\pi^{0}}$ and $R$ in the eight data periods.}
\label{RESULTS}
\end{table}

The constraint on the total momentum of photons in the kinematic fit is very effective in reducing the error on
 the two-photon invariant mass. The absolute scale of $W$ is  determined using  the CMD2 $m_{\phi}$ value as in the
$K$ mass measurement section.

Rescaling the values of Table~\ref{RESULTS} for the ratio
$m_{\phi}^{CMD2} / m_{\phi}^{KLOE}$ we obtain
\begin{equation}
   m_{\pi^{0}} = (134.906 \pm 0.012_{\rm stat} \pm 0.048_{\rm syst}) \quad
   \mbox{MeV}
   \label{EQ2}
\end{equation}
\begin{equation}
   m_{\eta} = (547.874 \pm 0.007_{\rm stat} \pm 0.031_{\rm syst}) \quad
   \mbox{MeV}
   \label{EQ3}
\end{equation}
the  $\pi^{0}$ mass value is in agreement with the world
average~\cite{PDG06} within 1.4$\sigma$. 

As a check of this result, we can use the measured ratio
\begin{equation}
   R = \frac{m_{\eta}}{m_{\pi^{0}}} = 4.0610 \pm 0.0004_{\rm stat} \pm 0.0014_{\rm syst}
   \label{EQ4}
\end{equation}
and the world average value of the $\pi^{0}$ mass, $m_{\pi^{0}} =
(134.9766 \pm 0.0006)$ MeV~\cite{PDG06} to derive
$m_{\eta} = (548.14 \pm 0.05_{\mathrm{stat}} \pm 0.19_{\rm syst})$ MeV,
consistent with the results quoted above although affected by a
larger systematic error due to a worse cluster position reconstruction
of the two photons from  $\pi^0$ decays which have a lower energy.
\section{Conclusions}

Our $K^0$ mass measurement is in acceptable agreement with  the previous measurements shown in Table~\ref{MASSK0}, but more accurate.  Averaging \cite{k0CMD2,k0NA48} and our result we obtain $M_{K^0} = 497.610 \pm 0.015$ MeV.
\begin{table}[hbt]
   \begin{center}
   \begin{tabular}{|lc|c|c|c|} \hline
   Experiment & & Method & $m_{K^0}$ (MeV) & events \\
   \hline
   CMD & \cite{k0CMD2} & $e^+ e^- \to K_L K_S$ & 
   497.661 $\pm$ 0.033  & 3713 \\
   NA48 & \cite{k0NA48} & $K_L \to 3 \pi^0 $&
   497.625 $\pm$ 0.001 $\pm$ 0.031 & 665 k \\
   KLOE & & $e^+ e^- \to K_L K_S$ &
   497.583 $\pm$ 0.005 $\pm$ 0.020 & 35 k\\
   \hline
   \end{tabular}
   \end{center}
   \caption{Recent measurements of the $K^0$ mass.} \label{MASSK0}
\end{table}

Our measurement of the $\eta$ (eq. \ref{EQ3}) mass is the most accurate result today. It is in good agreement 
with the recent measurements based on $\eta$
decays listed in Table~\ref{MASS}. Averaging the mass values from \cite{etamassNA48,etamassCLEO} and our result we
obtain $m_{\eta} = 547.851 \pm 0.025$ MeV, a value
different by $\sim$10$\sigma$ from the average of the measurements
done studying the production of the $\eta$ meson at threshold in
nuclear reactions.

\begin{table}[hbt]
   \begin{center}
   \begin{tabular}{|lc|c|c|} \hline
   Experiment & & Method & $m_{\eta}$ (MeV) \\
   \hline
   GEM, MM & \cite{etamassGEM} & $p\ d \to\ ^{3}He\ \eta$  &
   547.311 $\pm$ 0.028 $\pm$ 0.032  \\
   NA48, IM & \cite{etamassNA48} & $\eta \to 3\pi^{0}$ &
   547.843 $\pm$ 0.030 $\pm$ 0.041  \\
   CLEO-c, IM & \cite{etamassCLEO} &
   $\eta \to \gamma \gamma, 3\pi^{0}, \pi^{+} \pi^{-} \pi^{0}$
   & 547.785 $\pm$ 0.017 $\pm$ 0.057  \\
   KLOE, IM & & $\eta \to \gamma \gamma$  &
   547.874 $\pm$ 0.007 $\pm$ 0.031 \\
   \hline
   \end{tabular}
   \end{center}
   \caption{Recent measurement of the $\eta$-meson mass. IM stands for invariant mass of decay products, MM for missing mass at production.} \label{MASS}
\end{table}

\section{Acknowledgements}

We thank the DAFNE team for their efforts in maintaining low background running 
conditions and their collaboration during all data-taking. 
We want to thank our technical staff: 
G.F.~Fortugno and F.~Sborzacchi for their dedicated work to ensure an
efficient operation of 
the KLOE Computing Center; 
M.~Anelli for his continuous support to the gas system and the safety of
the
detector; 
A.~Balla, M.~Gatta, G.~Corradi and G.~Papalino for the maintenance of the
electronics;
M.~Santoni, G.~Paoluzzi and R.~Rosellini for the general support to the
detector; 
C.~Piscitelli for his help during major maintenance periods.
This work was supported in part
by EURODAPHNE, contract FMRX-CT98-0169; 
by the German Federal Ministry of Education and Research (BMBF) contract 06-KA-957; 
by Graduiertenkolleg `H.E. Phys. and Part. Astrophys.' of Deutsche Forschungsgemeinschaft,
Contract No. GK 742; 
by INTAS, contracts 96-624, 99-37; 
and by the EU Integrated Infrastructure
Initiative HadronPhysics Project under contract number
RII3-CT-2004-506078.

\small{

}

\end{document}